\documentclass[letterpaper, 10 pt, conference]{ieeeconf}

\IEEEoverridecommandlockouts
\usepackage{cite}
\usepackage{amsmath,amssymb,amsfonts}
\usepackage[ruled,vlined, linesnumbered]{algorithm2e}
\usepackage{gensymb}
\usepackage{siunitx}
\usepackage{graphicx}
\usepackage{booktabs}
\usepackage{textcomp}
\usepackage{mathtools}
\usepackage{xcolor}
\usepackage{hyperref}
\usepackage{todonotes}
\usepackage{amsfonts}
\usepackage{amssymb}
\usepackage{amsmath}
\usepackage{hyperref}
\usepackage{xcolor}
\usepackage{user_macros}
\usepackage{soul}
\usepackage{diagbox}
\usepackage{multirow}
\usepackage[nolist,nohyperlinks]{acronym}
\def\BibTeX{{\rm B\kern-.05em{\sc i\kern-.025em b}\kern-.08em
    T\kern-.1667em\lower.7ex\hbox{E}\kern-.125emX}}
    
\newcommand{\sgnote}[1]%
{\textcolor{green}{\textbf{Note: #1}}}
\newcommand{\shnote}[1]%
{\textcolor{blue}{\textbf{Note: #1}}}

\usepackage{siunitx} 

\newcommand{\mwnote}[1]{}
\newcommand{\mknote}[1]{}

\begin{document}
\bstctlcite{IEEEexample:BSTcontrol}
\begin{acronym}
\acro{HJ}{Hamilton-Jacobi}
\acro{HJI}{Hamilton-Jacobi-Isaacs}
\acro{ODE}{Ordinary Differential Equation}
\acro{MPC}{Model Predictive Control}
\acro{MDP}{Markov Decision Processes}
\acro{RMSE}{Root Mean Squared Error}
\acro{MSE}[MSE]{mean-squared-error}
\acro{RL}{Reinforcement Learning}
\acro{PDE}{Partial Differential Equation}
\acro{USV}{Unmanned Surface Vehicle}
\acro{NGR}{Net Growth Rate}
\acro{DP}{Dynamic Programming}

\end{acronym}

\title{\Large \bf
Maximizing Seaweed Growth on Autonomous Farms:\\
A Dynamic Programming Approach for Underactuated Systems \\ Operating in Uncertain Ocean Currents}


\author{Matthias Killer$^{1,2,*}$, Marius Wiggert$^{1,*}$, Hanna Krasowski$^{1,2}$, Manan Doshi$^{3}$,\\  Pierre F.J. Lermusiaux$^{3}$ and  Claire J. Tomlin$^{1}$
\thanks{$^{*}$ M.K. and M.W. have contributed equally to this work. }
\thanks{$^{1}$ M.K., M.W., H.K., and C.J.T. are with the Department of Electrical Engineering and Computer Sciences, University of California, Berkeley, USA. For inquiries contact: {\tt\small mariuswiggert@berkeley.edu}}
\thanks{$^{2}$ M.K. and H.K. are with the 
School of Computation, Information and Technology of the 
Technical University of Munich, Germany}
\thanks{$^{3}$ M.D. and P.F.J.L. are with the 
Department of Mechanical Engineering at the 
Massachusetts Institute of Technology, USA.}
\thanks{The authors gratefully acknowledge the support of the C3.ai Digital Transformation Institute and 
the IFI program of the German Academic Exchange Service (DAAD).}
}

\maketitle


\begin{abstract}

Seaweed biomass presents a substantial opportunity for climate mitigation, yet to realize its potential, farming must be expanded to the vast open oceans. However, in the open ocean neither anchored farming nor floating farms with powerful engines are economically viable. Thus, a potential solution are farms that operate by \textit{going with the flow}, utilizing minimal propulsion to strategically leverage beneficial ocean currents.
In this work, we focus on low-power autonomous seaweed farms and design controllers that maximize seaweed growth by taking advantage of ocean currents. We first introduce a \ac{DP} formulation to solve for the growth-optimal value function when the true currents are known. However, in reality only short-term imperfect forecasts with increasing uncertainty are available. Hence, we present three additional extensions. Firstly, we use frequent replanning to mitigate forecast errors. 
Second, to optimize for long-term growth, we extend the value function beyond the forecast horizon by estimating the expected future growth based on seasonal average currents. Lastly, we introduce a discounted finite-time \ac{DP} formulation to account for the increasing uncertainty in future ocean current estimates. We empirically evaluate our approach with 30-day simulations of farms in realistic ocean conditions. Our method achieves 95.8\% of the best possible growth using only 5-day forecasts. \change{This demonstrates that low-power propulsion is a promising method to operate autonomous seaweed farms in real-world conditions.} 



\end{abstract}

\section{Introduction}
\label{sec:introduction}
Recent research has shown promising applications of seaweed biomass for climate mitigation. It can be used as human food, as cattle feed that reduces methane emissions \mwnote{from burping} \cite{roque2019effect}, for biofuel and plastic \mwnote{to replace oil} \cite{yong2022seaweed}, and for carbon capture i.e. when the biomass is sunk to the ocean floor, it removes carbon dioxide from the atmosphere \cite{drawdown}. To deliver on this promise, production must scale by expanding seaweed farming from labor-intensive operations near shore to automated solutions utilizing the vast expanse of the open oceans \cite{tullberg2022review}. But conventional farming becomes economically infeasible in deeper waters as anchoring costs increase with depth \cite{ross2022seaweed}.

\begin{figure}[ht!]
    \includegraphics[width=.48\textwidth,
    trim={0.1cm 0.1cm 0cm 0cm}
    ,clip]{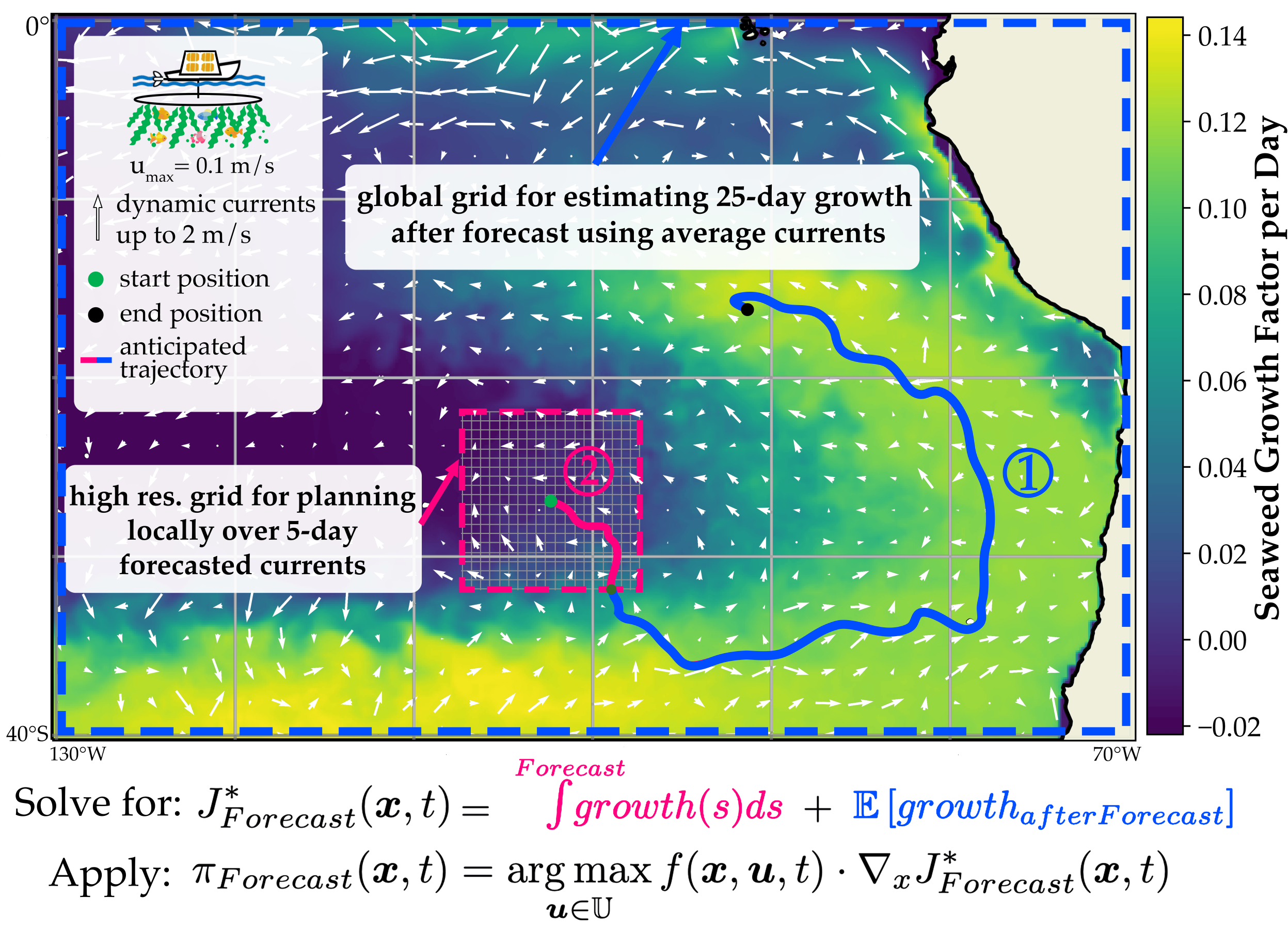}
    \caption{\small 
    Our method maximizes long-term growth on autonomous seaweed farm that operate by harnessing ocean currents. 
    We solve for the value function $J^*_{Forecast}$ that is long-term growth-optimal under the forecast with dynamic programming. We first compute the expected 25-day growth after the forecast based on historical average currents (\textbf{1}) and then use it to regularly solve for the value function over the next 5 days using daily current forecasts (\textbf{2}). Applying the induced policy $\pi_{Forecast}$ as feedback controller ensures high growth despite imperfect short-term forecasts. \mwnote{due to frequent replanning}
    }
    \label{fig:head_picture}
\end{figure}

A promising solution could be non-tethered, autonomous seaweed farms that roam the oceans while growing seaweed \cite{phykos, sherman2018seaweedpaddock}. These floating farms needs to be able to control their position to prevent stranding, colliding with ships, or drifting to nutrient-depleted waters. While they could be steered with powerful ship engines, the power and hence energy costs are prohibitively high due to the drag force scaling quadratically with the relative velocity of the farm. Recent studies \cite{wiggert.2022, doshi.2022} demonstrated that an autonomous vessel can navigate reliably to nearby targets by \textit{going with the flow}, using its minimal propulsion ($0.1 \frac{m}{s}$) strategically to nudge itself into ocean currents ($[0-2 \frac{m}{s}]$) that drift towards its destination. 
These studies have been extended to reduce the risk of stranding by incorporating obstacles \cite{doering_CDC2023} and to multi-agent fleets of vessels that navigate while staying connected in a local communication network \cite{hoischen_CDC2023}.
\change{In this paper, we use this low-power steering paradigm for operating seaweed farms. In contrast to \cite{wiggert.2022, doshi.2022}, which solves navigating to a target within a 5-day forecast horizon, our objective is to maximize seaweed growth along the trajectory of the farms over longer periods beyond the forecast horizon.}
For an autonomous vessel operating approach, there are four key challenges that we need to address. First, the currents are non-linear and time-varying. 
Second, in realistic settings, only coarse uncertain forecasts are available \cite{metzger2020validation, hycom, CopernicusGlobal,lermusiaux_et_al_O2006b,lermusiaux_JCP2006}. Third, the farm itself is \textit{underactuated} by which we mean that its propulsion is smaller than the surrounding currents, so it cannot easily compensate for forecast errors. 
Lastly, we want to maximize seaweed growth over weeks but forecasts from the leading providers are only 5-10 days long \cite{CopernicusGlobal, hycom} and \mwnote{similar to weather forecasts} the uncertainty for long-time predictions is high \mwnote{due to the chaotic system} \cite{pinardi_et_al_preface_TheSea2017,pinardi_et_al_TheSea2017}. \mwnote{Use the headfigure to explain why long-term reasoning vs greedy is important (i.e. a close medium growth field and a further away high growth field).} In a nutshell, we are tackling long-term horizon optimization of a state-dependent running cost with an underactuated agent in non-linear time-varying dynamics under uncertainty that increases over time. The long-term dependency of seaweed growth means the objective cannot easily be decomposed into multiple short-term objectives. While we showcase the method specifically on autonomous seaweed farms, there are many other potential applications of using the environmental dynamics such as navigating over long timeframes in the currents e.g. for low-powered sensor systems or on the winds with balloons, zeppelins for sensing our atmosphere or building mesh networks for telecommunication.


\subsection{Related Work}
Various approaches for time- and energy-optimal path planning exist for non-linear, time-varying dynamics like ocean currents \cite{lolla2012path, subramani_et_al_Oceans2017, lermusiaux_et_al_Oceanog2017, zhang2008optimal, tsou2013ant, huynh2015predictive, kularatne2018going, mannarini2019visir, kularatne2018optimal, subramani2019risk, pereira2013risk, hollinger2016learning, knizhnik2022flow, gunnarson2021learning, al2013wind, chakrabarty2013uav, bellemare2020autonomous}.
In the context of navigating within \textit{known currents or flows}, researchers have derived \ac{HJ} reachability equations for exact solutions \cite{lolla2012path, subramani_et_al_Oceans2017, lermusiaux_et_al_Oceanog2017}, non-linear programming \cite{zhang2008optimal, jones2017planning}, evolutionary algorithms \cite{tsou2013ant}, and graph-based search methods \cite{otte2016any, huynh2015predictive,chakrabarty2013uav, mannarini2019visir}. However, 
the last three 
techniques are prone to discretization errors and the non-convex nature of the problem, can lead to infeasibility or solvers getting stuck in local minima. 
In contrast, \ac{DP} based on the \ac{HJ} equations can solve the exact continuous-time control problem. \mwnote{Recent research has extended the \ac{HJ} framework \cite{lolla_et_al_OD2014_part1} to multi-time optimal navigation in underactuated current settings \cite{doshi.2022, wiggert.2022}.}

There is limited research that focuses on maximizing seaweed growth. 
In \cite{bhabra_et_al_Oceans2020, bhabra_MSThesis2021}, the authors maximize seaweed harvesting using autonomous vessels in varied settings. They use a 3D \ac{HJ} reachability framework in which the harvesting state is augmented into the third dimension. They optimize harvesting from stationary seaweed farms and assume the currents to be known making it not applicable in realistic settings, additionally the value function is not suitable for a closed-loop control policy.

For managing current uncertainty, previous work optimizes the expectation or a risk function over a stochastic solution of probabilistic ocean flows \cite{subramani_lermusiaux_CMAME2019}. However, this is not yet suitable for operational settings as it demands a principled uncertainty distribution for flows but most operational forecasts are deterministic. At the same time, robust control techniques, which aim to maximize the objective even in the face of worst-case disturbances, are not suitable when considering realistic error bounds, as the forecast error often equals or exceeds our low propulsion capabilities. 
Thus, to mitigate forecast inaccuracies, frequent replanning in a \ac{MPC} fashion has been proposed using either non-linear programming \cite{yilmaz_et_al_JOE2008,yilmaz_et_al_Oceans2006} or employing the \ac{HJ} value function as feedback policy \cite{wiggert.2022}, which offers the benefits of being both fast and optimal.
Another approach is to use \ac{RL} to learn how to best operate stratospheric balloons despite wind forecast uncertainty \cite{loon.2021,bellemare2020autonomous}. \change{While they ran operational experiments over more than 30 days their objective of staying above a certain area is relatively short-term, rendering \ac{RL} appropriate. However, the applicability of \ac{RL} for long-term objectives, similar to ours, remains uncertain.}

To address the increasing complexity associated with long-time horizons, problems are frequently divided into multiple subproblems using graph-based methods or hierarchical \ac{RL} \cite{eysenbach.2019.search.replay.buffer,pateria.2021.hierarchical.rl}. These approaches are appropriate for combinatorial optimization problems, where dividing and conquering in subtasks is effective. However, this is not suitable for our problem involving continuous state space and long-time dependencies.
A potential solution to handle growing uncertainty of the currents over time is to discount future rewards which is common in \ac{RL} settings \cite{sutton, doya2000reinforcement} and we do below.

\subsection{Overview of Method \& Contributions}
In this paper, we make five main contributions towards controllers that optimize seaweed growth on autonomous seaweed farms over long periods.

First, we formulate maximizing seaweed growth on an autonomous farm as an optimization problem that can be solved exactly with \ac{DP} in the 2D spatial state of the system (Sec. \ref{subsec:cost_derivation}). Compared to prior work using \ac{HJ} Reachability in 3D \cite{bhabra_et_al_Oceans2020} to model seaweed growth on stationary farms, our formulation leads to two advantages: significant reduction of computational complexity (Sec. \ref{sec:method}) and the value function can be used as feedback policy. This allows for frequent replanning in the \ac{MPC} spirit for multiple farms which is critical to mitigate forecast uncertainty \cite{hoischen_CDC2023}.
Second, we extend the value function beyond the forecast horizon which leads to a feedback policy that optimizes for long-term optimal growth (Sec. \ref{subsec:long_term_method}). 
\mwnote{For that we estimate the expected growth using historical average currents over a coarse grid and then initialize the \ac{DP} over the forecast horizon with these values (Fig. \ref{fig:head_picture}, Sec. \ref{subsec:long_term_method}).}
Third, to account for the growing uncertainty of the ocean current estimates, we introduce finite-time discounting into the \ac{DP} formulation
(Sec. \ref{subsec:discounting}).
Fourth, we are the first to run extensive empirical simulations of autonomous seaweed farms in realistic current settings over 30 days. We first investigate how different propulsion of the farms would affect the best achievable seaweed growth with known currents. We then evaluate how close different configurations of our method can get to the best achievable growth when only daily, 5-day forecasts are available (Sec. \ref{sec:experiments}). Lastly, we open-source our code, which contains extensive features to simulate, visualize, and study controllers for 2D vessels operating by harnessing uncertain ocean currents.

The remainder of the paper/article is structured as follows: in Sec. \ref{sec:ProblemStatement} we define the problem. Sec. \ref{sec:method} details the four components of our method. Sec. \ref{sec:experiments} contains the performance evaluation of our methods and baselines, and we conclude with Sec. \ref{sec:conclusion} and outline future work.

\section{Problem Statement}
\label{sec:ProblemStatement}

\subsection{System Dynamics}
We consider an autonomous seaweed farm as surface vessel on the ocean with the spatial state $\x \in \mathbb{R}^2$. Let the control input be denoted by $\bu$ from a bounded set $\mathbb{U} \in R^{n_u}$ where $n_u$ is the dimensionality of the control. Then, the spatial dynamics of the system at time $\t$ can be modelled by the first order \ac{ODE}:
\begin{equation}
\label{eq:spatial_ode}
    \dot{\x} \!=\! f(\x,\bu,\t) = v(\x, \t) + g(\x,\bu, \t), \quad \t \in [0, \T]
\end{equation}
where the movement of the vessel depends on the drift due to the time-varying, non-linear flow field $v(\x,\t) \to \mathbb{R}^2$ and its control $\bu$.
We choose a first-order model where the drift and control directly influence the state, disregarding inertial effects from motor acceleration and drag forces. This is justified by the fact that high-drag seaweed farms attain equilibrium velocity within a few minutes, a timescale considerably shorter than our 30-day planning horizon.

While our method is generally applicable, we focus on \textit{underactuated} settings in the sense that most of the time $\max\|g(\x,\bu, \t)\|_2 \ll \|v(\x, \t)\|_2$. \change{Note that our notion of underactuation differs from the common notion in control research, which refers to a system with fewer independent control actuators than degrees of freedom to be controlled.}
We denote the spatial trajectory induced by this \ac{ODE} with $\stt$. For a vessel starting at the initial state $\ist$ at time $\tzero$ with control sequence $\uall$, we denote the state at time $\t$ by $\ltraj{\t} \in \mathbb{R}^2$. The system dynamics (Eq. \ref{eq:spatial_ode}) are assumed to be continuous, bounded, and Lipschitz continuous in $\x, \bu$ \cite{doshi.2022}.

Additionally, we assume the farm has seaweed mass $\m$ which evolves according an exponential growth \ac{ODE}:
\begin{equation}
\label{eq:seaweed_ode}
    \dot{\m} \!=\! \m \cdot  \gf(\x, \t), \quad \t \in [0, \T]
\end{equation}
where $\gf$ is the growth factor per time unit, e.g. $20\frac{\%}{\text{day}}$, which depends on nutrients, incoming solar radiation, and water temperature at the spatial state $\x$ and time $\t$. 

\subsection{Problem Setting}
The objective of the seaweed farm starting from $\ist$ at $\tzero$ with seaweed mass $\m(\tzero)$ is to maximize the seaweed mass at the final time $\T$. This implies optimizing the growth along its trajectory $\trajraw$.
\begin{equation}
\label{eq:final_seaweed_eq}
\small
    \max_{\substack{\uall}} \; \m(\T) = \m(\tzero) + \max_{\substack{\uall}} \; \int_{\tzero}^\T \m(\s) \cdot 
    \underbrace{\gf (\trajInit{\s}, \s)}_{\substack{\text{growth factor}}}\;d\s
\end{equation}


If the currents $\trueCurrents$ are known, our method (Sec. \ref{sec:method}) is guaranteed to find the optimal value function from which the optimal control $\uoptall$ and trajectory can be obtained.
However, in realistic scenarios only inaccurate, short-term forecasts $\FCCurrents$ are available at regular intervals. These differ from the true flow $\trueCurrents$ by the forecast error $\delta(\x,\t)$.
Our goal is then to determine a feedback policy $\pi(\x,\t)$ that results in a high expected seaweed mass $\mathbb{E}[\m(\T)]$. Hence, in our experiments (Sec. \ref{sec:experiments}) we evaluate our method empirically over a set of missions $(\ist, \t) \sim \mathbb{M}$ and a realistic distribution of true and forecasted ocean currents $\trueCurrents, \FCCurrents \sim \mathbb{V}$.

\section{Method}
\label{sec:method}
Our method consists of a core \ac{DP} formulation that optimizes seaweed growth when the currents are known and three extensions to get a feedback policy $\pi$ that performs well over long-time horizons when only limited forecasts are available. We first introduce the core \ac{DP} formulation to obtain the growth-optimal value function (Sec. \ref{subsec:mpc_replanning}). Then, we demonstrate using the value function as feedback policy $\pi$, which is equivalent to replanning at every time step (Sec. \ref{subsec:long_term_method}). This leads to reliable performance even if the value function was computed with inaccurate forecasts.
Next, we extend the feedback policy by estimating the growth beyond the forecast horizon (Sec. \ref{subsec:long_term_method} and introduce a finite-time discount factor \ref{subsec:discounting}). Lastly, we describe the control algorithm variations developed and discuss computational aspects (Sec. \ref{subsec:controllers} and \ref{subsec:complexity}).

\subsection{Maximizing Seaweed Mass With Known Dynamics}
\label{subsec:cost_derivation}
We use continuous-time optimal control where the value function $\valfunc$ of a trajectory $\stt$ is based on a state and time-dependent reward $\rng$ and a terminal reward $\trm$: 
\begin{align}
\nonumber
\valfunc = \int_{t}^{\T} \rng(\traj{\s},\s)d\s + \trm(\traj{\T}, \T).
\end{align}
Let $\optvalfunc = \max_{\substack{\uall}} \valfunc$ be the optimal value function. Using \ac{DP} we can derive the corresponding \acl{HJ} \ac{PDE} \cite{bansal.2017}:

\begin{align}
\small
    \label{eq:hjb}
   -\pfracp{\optvalfunc}{\t} &= 
   \max_{\bu}\left[\Dx{\optvalfunc} 
    \cdot f(\x,\bu,\t) + \rng(\x, \t)
    \right] \\
    \label{eq:hjb_term}
    \svalfunc(\x,\T) &= \trm(\x, \T)
    .
\end{align}

\change{Computationally, this \ac{PDE} is solved in the state space discretized into N grid points along each dimension d \cite{mitchell2005toolbox}. At each step backward in time we need to compute the gradient $\Dx{\optvalfunc}$ which is $O(d)$ for each grid point, so the complexity scales exponentially with the state dimension as $O(dN^d)$, which is called the curse of dimensionality \cite{bansal.2017}.}

Next, we define the reward $\rng$ and terminal reward $\trm$ to maximize $\m(\T)$. One approach is to solve the \ac{PDE} in an augmented state space $x_{aug} \!=\! (\x, \m)^\top \in \mathbb{R}^3$. If we set $\trm\!=\!0$ and define the reward as $\rng\!=\!\m \cdot \gf(\x, \t)$, the value function is our objective (Eq. \ref{eq:final_seaweed_eq}). \change{However, as the computational complexity of solving for $\svalfunc$ scales exponentially with the state dimension, we want a reward $\rng$ that does not depend on the augmented state $\m$.}
 For that, we introduce the variable $\eta = \ln(\m)$ with the new dynamics $\dot{\eta} = \frac{\dot{\m}}{\m} =  \gf(\x, \t)$. As $\eta(\m)$ is strictly increasing in $\m$, the control $\uoptall$ that maximizes $\eta(\T)$ is equivalent to $\uoptall$ maximizing $\m(\T)$. We can then reformulate Eq. \ref{eq:final_seaweed_eq} to $\eta(\T)$:
 \begin{gather}
\small
    \label{eq:ln_new_opt}
    \max_{\substack{\uall}} \; \eta(\T) = \eta(\tzero) + \max_{\substack{\uall}} \int_{\tzero}^\T 
    \gf (\traj{\s}, \s)\;d\s .
\end{gather}
By setting the reward to $\rng = \gf(\x, \t)$ the optimal value function captures this optimization without requiring $\m$:
\begin{equation}
\label{eq:value_func}
\optvalfunc = \max_{\substack{\uall}} \int_{\t}^\T \gf (\traj{\s}, \s)\;d\s.
 \end{equation}

\change{We then solve the \ac{HJ} \ac{PDE} for the growth-optimal $\optvalfunc$ in the spatial state $\x$ and obtain $\uoptall$ and the trajectory $\opttrajraw$ that maximize $\m(\T)$ at $\frac{2}{3}N$ the computational cost (Sec. \ref{subsec:complexity}). This formulation can be applied more generally to optimize the state of exponential growth or decay \ac{ODE}s.} We can convert the value of $\valfuncInit$ to the final seaweed mass of the optimal trajectory starting at $\ist, \tzero$ with $\m(\tzero)$:
\begin{equation}
\nonumber
\m(\T) = \m(\tzero) \cdot e^{\int_{\tzero}^\T \gf (\opttraj{\s}, \s)\;d\s} = \m(\tzero) \cdot e^{\valfuncInit}.
\end{equation}



\subsection{Feedback Policy Based on Regular Forecasts}
\label{subsec:mpc_replanning}

The value function $\svalfunc$ from Sec. \ref{subsec:cost_derivation} allows us to compute the optimal control $\bu^*(\x,\t)$ for all $\x, \t$ and hence a feedback policy $\pi(\x,\t)$ for the vessel or multiple vessels in the same region \cite{wiggert.2022}. This policy is the optimizer of the Hamiltonian (right side Eq. \ref{eq:hjb}):
\begin{align}
\label{eq:policy}
\pi(\x,\t) &= \argmax_{\bu \in \mathbb{U}} f(\x, \bu, \t) \cdot \nabla_x \optvalfunc, 
\end{align}
which can often be computed analytically depending on $g(\x, \bu, \t)$. While $\pi$ is optimal if $\svalfunc$ is based on the true currents $\trueCurrents$, it can also be applied when imperfect forecasts $\FCCurrents$ were used to compute the value function $\valfuncFC$.
In that case, an agent at state $\x$ executing $\pi_{\FCCurrents}(\x,\t)$ will find itself at a different state $\x^\prime$ than anticipated as $\trueCurrents$ differs from $\FCCurrents$. But the control that would be growth optimal under $\FCCurrents$ can again be computed with $\pi_{\FCCurrents}(\x^\prime,\t + \Delta \t)$. Applying $\pi_{\FCCurrents}$ closed-loop is hence equivalent to full-time horizon re-planning with $\FCCurrents$ at each time step. \change{This notion of re-planning at every time step at the low cost of a 2D gradient computation (Sec. \ref{subsec:complexity}) ensures good performance despite forecast errors \cite{wiggert.2022}.} $\valfuncFC$ can be updated daily as new forecasts arrive.


\subsection{Reasoning Beyond the Forecast Horizon}
\label{subsec:long_term_method}
As the growth cycles of seaweed typically spans months, our aim is to maximize the seaweed mass at an \textit{extended} future time $\Text$ after the final time of the 5-day forecast $\Tfc$. A principled way to reason beyond the planning horizon is to estimate the expected growth our seaweed farm will experience from the state $\traj{\Tfc}$ onward and add this as terminal reward $\trm$ to Eq. \ref{eq:value_func}.
\begin{multline}
\label{eq:value_func_with_terminal_reward}
\small
    \valfuncExt = \valfuncFCTfc + \mathbb{E}\left[J^*_{\Text}(\traj{\Tfc},\Tfc) \right] \\
    \valfuncFCTfc  = \max_{\substack{\uall}} \int_{t}^{\Tfc} \!\!\!\!\!\!\!\! \gf (\traj{\s}, \s)\;d\s
\end{multline}
where $\valfuncFCTfc$ is the growth a vessel starting from $\x$ at $\t$ will achieve at $\Tfc$ and $\small \mathbb{E}\left[J^*_{\Text}(\traj{\Tfc},\Tfc) \right]$ estimates the additional growth from $\Tfc$ to $\Text$. The expectation is over the uncertain future ocean currents. 

We propose to estimate $\mathbb{E}\left[J^*_{\Text} \right]$ by computing the value function $J^*_{\AvgCurrents,\Text}$ based on monthly average currents $\AvgCurrents$ for the region using Sec. \ref{subsec:cost_derivation}. To compute $\svalfuncExt$ we then solve Eq. \ref{eq:hjb} with $\trm(\x, \Tfc)=J^*_{\AvgCurrents,\Text}(\x, \Tfc)$.


\subsection{Finite-time Discounting to Mitigate Uncertainty}
\label{subsec:discounting}
As the oceans are a chaotic system, the uncertainty of the forecasted ocean currents increases over time. We can incorporate this increasing uncertainty in the value function by using the finite-time discounted optimal control formulation:
\begin{align}
\small
\nonumber
\valfuncdisc = \int_{t}^{\T} \!\!\!\! e^{\frac{-(s-t)}{\tau}} \rng(\traj{\s},\s) \;d\s + \trm(\traj{\T}, \T),
\end{align}
where $\tau$ is the discount factor.
\change{Note that in contrast to discrete time dynamics, where discount factors range from 0 to 1, our application of  $\tau$ conforms to the conventional interpretation in continuous dynamic programming \cite{doya2000reinforcement}: $\tau$ can assume values significantly greater than 1 and for higher $\tau$ future rewards are discounted less.}
We derive the corresponding \ac{HJ} \ac{PDE} by following the steps in \cite{doya2000reinforcement} and in place of Eq. \ref{eq:hjb} we obtain:
\begin{equation}
\small
\nonumber
   \pfracp{\valfuncoptdisc}{\t} = -
   \max_{\bu}\left[\Dx{\svalfuncoptdisc} 
    \cdot f(\x,\bu,\t) + \rng (\x, \t)
    \right] + \frac{\valfuncoptdisc}{\tau}
\end{equation}

\subsection{Control Algorithm Variations}\label{subsec:controllers}


All variations of our method are feedback policies $\pi$ derived from a value function (Sec.\ \ref{subsec:mpc_replanning}). The four variations differ only in how the value function is computed. When the true currents $\trueCurrents$ are known we compute $\svalfunc$ (Eq. \ref{eq:value_func}) for optimal control. 
When only forecasts $\FCCurrents$ are available, we calculate the $\svalfuncFC$ for planning horizons up to the end of the forecasts $\Tfc$ and update it as new forecasts become available (Sec. \ref{subsec:mpc_replanning}). Thirdly, to optimize for growth until $\Text > \Tfc$ we calculate an extended value function $\svalfuncExt$ (Sec. \ref{subsec:long_term_method}) using average currents $(\FCCurrents + \AvgCurrents)$. Lastly, we can discount future rewards with $\svalfuncoptdisc$ (Sec. \ref{subsec:discounting}) in any of the above value functions. In Algorithm \ref{alg:closed_loop} we detail the discounted, long-term version as it contains all components.

\setlength{\textfloatsep}{15pt}
\begin{algorithm}
\vspace{1mm}
\caption{Discounted \ac{HJ} Closed-loop Control}
\label{alg:closed_loop}
\small
\KwIn{\text{Forecast Flow(s)} $\FCCurrents$, $t=0$, $\vect{x}(t)=\vect{x}_0$, average Flows $\AvgCurrents$, discount $\tau$, plan until $\Text$}
Compute $J^{*, \tau}_{\AvgCurrents,\Text}$ using $\bar{v}$ (Sec. \ref{subsec:long_term_method})\;
\While{$t \leq T$}{
  \If{new forecast $\FCCurrents$ available}{
    Compute $J^{*, \tau}_{\FCCurrents, \text{ext}}$ (Sec. \ref{subsec:long_term_method})\;
  } 
  $\bu_t = \pi^{*, \tau}_{\FCCurrents, \text{ext}}(x_t,t)$; using $J^{*, \tau}_{\FCCurrents, \text{ext}}$  (Sec. \ref{subsec:mpc_replanning}) \\ 
  
  $x(t+\Delta t) = x(t) + \int_{t}^{t+\Delta t} f(\bu_t,x(s),s) \,ds $\;
  $t \gets t+\Delta t$\;
}
\vspace{-2mm}
\end{algorithm}


\change{\subsection{Computational Considerations}
\label{subsec:complexity}
To illustrate the computational advantage of our approach let’s consider our realistic simulation experiments in Sec. \ref{sec:experiments}. The computational complexity is $O(dN^d)$ so with a spatial discretization of $N=120$ solving for $\svalfunc$ in $d=3$ compared to $d=2$ dimensions would be $\frac{2}{3}N=180$ times as expensive. We only need to solve the 2D \ac{HJ} \ac{PDE} for $\valfuncFC$ once per day as new forecasts become available. From the value function, we obtain the optimal control every 10 minutes with just a cheap 2D gradient computation $O(d)$. In contrast, using non-linear programming \ac{MPC}, we would need to solve an optimization problem 144 times per day. Additionally, non-linear programming \ac{MPC} does not provide convergence and optimality guarantees, which are provided for our solution due to solving the 2D \ac{HJ} \ac{PDE}. Moreover, $\valfuncFC$ can be used for hundreds of farms in the same region \cite{hoischen_CDC2023}, whereas \ac{MPC} would need to be run for each farms.

One limitation of \ac{DP} with ocean current forecasts provided as matrices is that it requires significant RAM due to the interpolation of currents at each time-step. We use JAX to first compile a computational graph for the value function computation before solving the \ac{PDE}. This adaption yields a significant speed-up over the Matlab-based helperOC. Nevertheless, it took 60GB of RAM for 30 day planning of $\valfuncExt$, which limited our simulation horizon. This can be optimized further e.g. by using GPU acceleration and moving the interpolation outside of the \ac{PDE} solving.
}
\section{Experiments}
\label{sec:experiments}
As our system is underactuated (Sec. \ref{sec:ProblemStatement}), it is impossible to prove robustness of our method against potentially adversarial currents \cite{doering_CDC2023}. Hence, we evaluate our method empirically by simulating the operation of an autonomous seaweed farm in realistic ocean currents and growth conditions. We will open-source the code for our simulator and controllers for others to replicate results and build on\footnote{The code will be available in a github repository
}.
We run two main experiments.
First, we investigate how varying the propulsion $\bu_{max}$ impacts the best achievable seaweed growth under known currents $\trueCurrents$ and compare it to the growth achieved by 30-day planning without discounting relying on daily, 5-day forecasts $\trueCurrents$ and average currents $\AvgCurrents$ (Sec. \ref{subsubsec:exp_u_max}). 
Second, we fix the propulsion to $\bu_{max}\!\!=\!\!0.1\frac{m}{s}$ and evaluate how the planning horizon and discounting in our method affect growth and how close we can get to the best achievable growth while relying on daily forecasts $\FCCurrents$ and average currents $\AvgCurrents$ (Sec. \ref{subsubsec:exp_u_max_fixed}). The experimental setup for both is the same and will be explained next.

\subsection{Experimental Setup}
\label{subsec:experimental_setup}
\subsubsection{Seaweed Growth Model}
Macroalgae growth depends on the species, the water temperature, solar irradiance, and dissolved nutrient concentrations, specifically nitrate $(\mathrm{NO}_3)$ and phosphate $(\mathrm{PO}_4)$ \cite{bhabra_et_al_Oceans2020}. We use the model of the \ac{NGR} of Wu et al. \cite{Wu2022.seaweedModel}\mwnote{as it models the key growth dependencies without maintaining additional state variables to model the plant-internal nutrient state \cite{MARTINS2002247}} and temperate species parameters from \cite{MARTINS2002247, zhang.2016.model}. In this model, the time-dependent \ac{NGR} is determined by the growth rate $r_{growth}$ and the respiration rate $r_{resp}$ caused by metabolism as:
\begin{equation}
\label{eq:d_biomass}
    \dot{\m}(t) \!=\! \m(t) \cdot  \mathrm{NGR}(t) \!=\! \m(t) \cdot (r_{growth}(t) -r_{resp}(t)) .
\end{equation}
Fig.\;\ref{fig:head_picture} shows the \ac{NGR} for our region at the apex of the sun's motion in January 2022. 



\subsubsection{Realistic Ocean Forecast Simulation}
In realistic operations the vessel receives daily forecasts for replanning. In our simulations, we use Copernicus \cite{CopernicusGlobal} hindcasts as true currents $\trueCurrents$ and mimic daily 5-day forecasts $\FCCurrents$ by giving the planner access to a 5-day sliding time window of HYCOM \cite{hycom} hindcasts. Aligned with previous work \cite{wiggert.2022}, we find that the forecast error $\delta$ with this setting is comparable to the evaluated forecast error of HYCOM \cite{metzger2020validation} in key metrics.
To estimate the expected growth beyond the forecast horizon of $\FCCurrents$ (Sec. \ref{subsec:long_term_method}) we use $\frac{1}{6}\text{th} \deg$ seasonal averages $\AvgCurrents$ of the ocean currents from Copernicus 2021. 
\subsubsection{Large Scale Mission Generation}
\label{subsubsec:miss_gen}
We simulate operations in the southeast Pacific due to high nutrient densities.
For a large representative set of missions $\mathbb{M}$, we sampled 1325 tuples ($\ist, \tzero, \m(\tzero)\!\!=\!\!\text{100kg}$), uniformly distributed in time between January and October 2022 and across the region of longitude [-130, -70]$\degree$W and latitude [-40, 0]$\degree$S. This allows for varying current distributions.\mwnote{The samples were generated maintaining a distance of 0.5 degrees from land to avoid any instant collisions.} 
As our method is not aware of land obstacles, we had 290 missions where at least one of the farms stranded or left the simulation region. While stranding can be avoided by modifying the \ac{HJ} \ac{PDE} \cite{doering_CDC2023}, we consider only the remaining 1035 missions for our results.
\mwnote{Each mission starts with a seaweed mass of 100kg.}

\subsubsection{Evaluated Controllers and Baselines}
We evaluate our method in different configurations categorized by: a) the ocean current data used by the controller for planning, either the true currents $\trueCurrents$ or daily forecasts $\FCCurrents$ and average currents $\AvgCurrents$, and b) the controller's planning horizon $\Text$ over which it optimizes growth, either the entire 30-day period or more short term \textit{greedy} (5-day and 1 hour). We also examine the effect of using of a discounted value function. An overview of the configurations is provided in Tab. \ref{tab:settings}.

The simplest baseline that we compare against is the seaweed growth on passively floating farms. We also consider the greedy \ac{HJ}-based controllers as baselines representing all short-term controllers. That is because they are optimal under $\FCCurrents$, hence we would only expect \ac{MPC} or another approach to be better in the unlikely case that their approximation errors would systematically improve performance.

For long-term ($\Text$=30 days) controllers, we compute the growth-to-go after $\Tfc$, i $J^*_{\AvgCurrents,\Text}(\x, \Tfc)$,  over the full area on a coarse $\frac{1}{6}\degree$ grid, as illustrated in Fig. \ref{fig:head_picture}. The value function $\valfuncExt$ used for the control policy is then computed daily on new forecasts using a smaller $\frac{1}{12}\degree$ grid around the current farm's position (10$\degree$ square).

\subsubsection{Evaluation Metrics}
Our objective is to maximize the seaweed mass at the end of each mission $\m(\T)$. Additionally, we compute the relative improvement in final seaweed mass by normalizing within each mission with the baseline final mass. We then present the average relative improvement across all missions which allows us to gauge how much more/less biomass a specific controller can grow above the baseline. This is important as the start $\x_0$ of a mission is a major indicator of achievable growth as illustrated in Fig.~\ref{fig:heatmap}. As baselines we use either passively floating or the best achievable growth based on the true currents $\trueCurrents$.

\begin{table}[t]\centering 
\vspace{3mm}
\caption{Compared controller settings}
\label{tab:settings}
\begin{tabular}{ccc} \toprule
\textbf{controller}       & \textbf{planning horizon $\Text$} & \textbf{discount $\tau$} \\ \hline
w/o discount ($\trueCurrents$) & 30 days                   & -                            \\
floating                         & -                         & -                            \\
greedy 1 hour ($\FCCurrents$)                    & 1 hour                   & -                            \\
greedy 5 days ($\FCCurrents$)                   & 5 days                    & -                            \\
w/o discount ($\FCCurrents$ + $\AvgCurrents$)    & 30 days                   & -                            \\
w/ discount I ($\FCCurrents$ + $\AvgCurrents$)  & 30 days                   & 1.296.000                    \\
w/ discount II ($\FCCurrents$ + $\AvgCurrents$)  & 30 days                   & 1.728.000                    \\ \hline
\end{tabular}
\vspace{-5mm}
\end{table}



\subsection{Experimental Results}
\label{subsec:experimental_results}

\subsubsection{How does varying propulsion affect growth?}
\label{subsubsec:exp_u_max}
\begin{figure}[t]
    \includegraphics[width=.44\textwidth,
    ,clip]{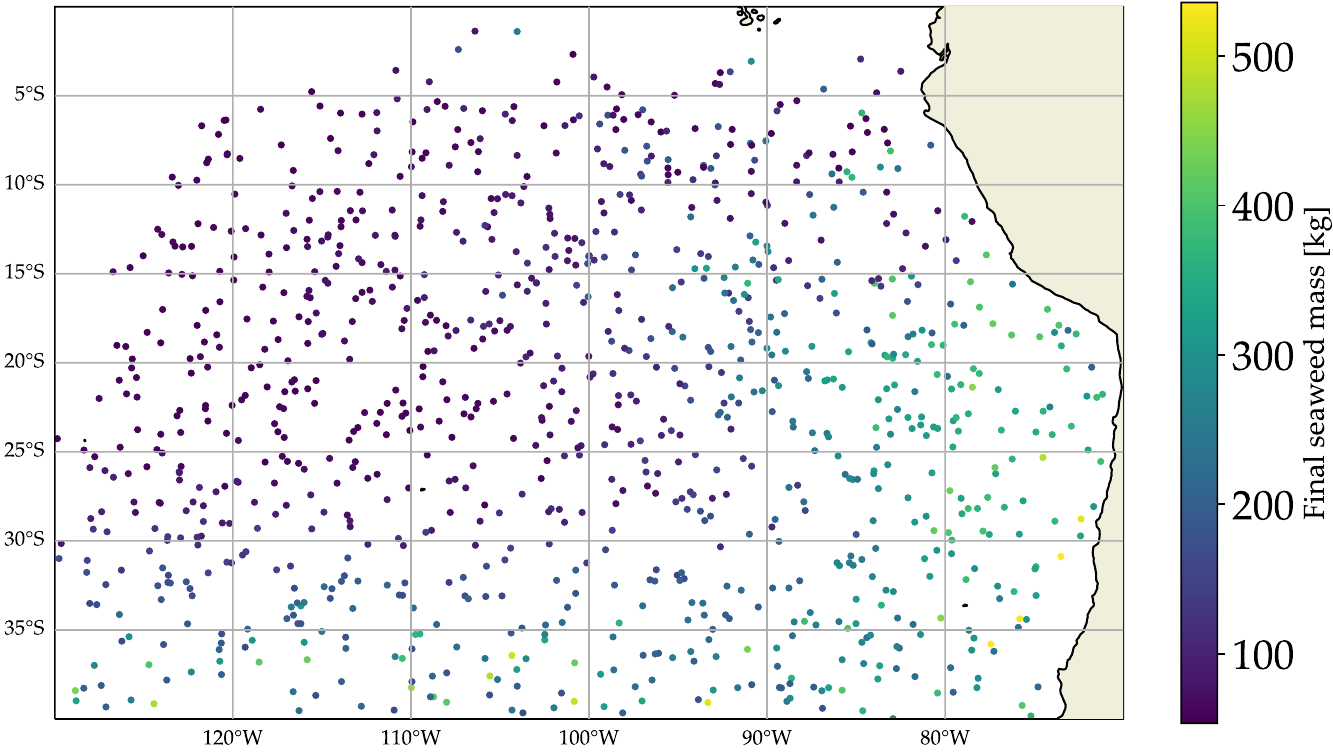}
    \caption{\small We sample a diverse set of starts ($\x_0, \tzero$) for seaweed farms to empirically evaluate our controllers. The colorized starts show the best achievable seaweed mass after 30 days using $u_{max}=0.1\frac{m}{s}$.}
    \label{fig:heatmap}
    \vspace{-1mm}
\end{figure}

We vary the maximum propulsion $\bu_{max}$ of the farm and evaluate how this impacts the best achievable seaweed growth under known currents $\trueCurrents$. Fig. \ref{fig:box_plot_umax} and Tab. \ref{tab:normalized_performance_umax} compare the final seaweed mass distributions for different propulsion levels, starting with passively floating. We observe that the average seaweed growth scales almost linearly with $\bu_{max}$, yielding between 15\% and 12\% more biomass per $0.1\frac{m}{s}$ propulsion. We also compare how much growth our method w/o discount ($\FCCurrents$ + $\AvgCurrents$) achieves with varying propulsion. As expected this achieves slightly less biomass ($\approx$95-96\% of $\trueCurrents$) due to forecast errors for all propulsion levels. For higher $\bu_{max}$ the gap is slightly smaller, possibly because the farm can better compensate for forecast errors. Nonetheless, even small propulsion of $u_{max}\!\!=\!\!0.1\frac{m}{s}$ enables 9.6\% more biomass than a passively floating farm.
The start $\x_0$ of a mission significantly influences 30-day growth, as shown in Fig. \ref{fig:heatmap}. High-growth missions are situated in the east and south of our region, aligning with nutrient-rich areas (see Fig. \ref{fig:head_picture}).


\begin{table}[b] \centering
\caption{Average seaweed growth for different propulsions $\bu_{max}$}
\vspace{-2mm}
\label{tab:normalized_performance_umax} 
\begin{tabular}{@{}cccc@{}}
\toprule
\textbf{$u_{max}$} & \textbf{planning input} & \textbf{relative growth} & \textbf{final seaweed mass} \\ \midrule
0.0$\frac{m}{s}$                  & (floating)            & 100\%             & 145.29kg$\pm$100.30kg                     \\
\multirow{2}{*}{0.1$\frac{m}{s}$}      & $\trueCurrents$             & 115.38\%   & 166.45kg$\pm$109.67kg             \\
                                & $\FCCurrents$ + $\AvgCurrents$   & 109.62\%     & 159.29kg$\pm$107.46kg  \\
\multirow{2}{*}{0.2$\frac{m}{s}$}      & $\trueCurrents$             & 128.69\%   & 182.04kg$\pm$115.11kg       \\
                          & $\FCCurrents$ + $\AvgCurrents$         & 121.29\%    & 173.72kg$\pm$112.94      \\
\multirow{2}{*}{0.3$\frac{m}{s}$}      & $\trueCurrents$             & 141.27\%  & 194.98kg$\pm$117.39kg      \\
                          & $\FCCurrents$ + $\AvgCurrents$          & 133.28\%   & 187.01kg$\pm$116.60kg    \\
\multirow{2}{*}{0.4$\frac{m}{s}$}    & $\trueCurrents$             & 153.71\%    & 206.96kg$\pm$118.34kg     \\
                          & $\FCCurrents$ + $\AvgCurrents$       & 145.79\%     & 199.50kg$\pm$118.09kg     \\
\multirow{2}{*}{0.5$\frac{m}{s}$}      & $\trueCurrents$             & 165.79\%  & 218.10kg$\pm$118.59kg  \\
                          & $\FCCurrents$ + $\AvgCurrents$       & 158.14\%   & 210.78kg$\pm$117.72kg                     \\ \bottomrule
\end{tabular}
\vspace{1mm}
\label{tab:normalized_performance_umax} 
{\footnotesize Planning on $\trueCurrents$ is the best achievable which we compare to using forecasted and average currents ($\FCCurrents$ + $\AvgCurrents$) without discounting. Relative growth is normalized per mission by passively floating.}
\vspace{1mm}
\end{table} 

\begin{figure}[t]
    \vspace{2mm}
    \includegraphics[width=.48\textwidth,]{figures/seaweed_vs_umax_final_adapted-2_compr.png}
    \caption{\small 
    The best achievable seaweed mass given $\trueCurrents$ increases linearly with $u_{max}$. Operating with our long-term control method using forecasts $\FCCurrents$ and average currents $\AvgCurrents$ achieves $\approx$ 95\% of growth.
    }
    \label{fig:box_plot_umax}
\end{figure}

\subsubsection{The impact of planning horizon and discounting}
\label{subsubsec:exp_u_max_fixed}
As the energy consumption scales cubically with $\bu_{max}$, higher propulsion may be economically infeasible for real-world applications. Therefore, for this experiment we fix $\bu_{max}\!\!=\!\!0.1\frac{m}{s}$. We investigate how different planning horizons and discounting affect performance when operating with forecasts $\FCCurrents$ and how close we can get to the best achievable growth. We evaluate two greedy controllers that repeatedly optimize over short $\Text$ (1h and 5-days) and compare to 30-day time-horizon with different discounting settings (Tab. \ref{tab:settings}).

Table \ref{tab:normalized_performance} shows the results. As expected, both the greedy and long-term controllers outperform passively floating. Surprisingly, the performance of the 5-day greedy controller, is close to the 30-day controllers. Using the discounted formulation slightly improves the long-term controller, yielding on average 95.77\% of the best achievable growth. 

\begin{table}[b]\centering
\caption{
 Seaweed growth of different controllers for 1035 missions}
\vspace{-2mm}
\label{tab:normalized_performance}
\begin{tabular}{@{}ccc@{}}

\toprule
\textbf{controller $\bu_{max}\!=\!0.1\frac{m}{s}$}  & \textbf{relative growth}  & \textbf{final seaweed mass} \\ \midrule
w/o discount ($\trueCurrents$)  & 100\%         & 168.45kg$\pm$109.67kg                         \\
floating (-)          & 88.20\%       & 145.29kg$\pm$99.54kg                         \\
greedy 1 hour ($\FCCurrents$) & 92.24\%       & 152.48kg$\pm$102.89kg                       \\
greedy 5 days ($\FCCurrents$)  & 95.19\%       & 157.78kg$\pm$106.04kg                         \\
w/o discount ($\FCCurrents$ + $\AvgCurrents$)   & 95.61\%       & 158.84kg$\pm$106.71kg                         \\
w/ discount I ($\FCCurrents$ + $\AvgCurrents$)  & 95.77\%       & 159.16kg$\pm$106.62kg                         \\
\textbf{w/ discount II ($\FCCurrents$ + $\AvgCurrents$)}  & \textbf{95.77}\%        & \textbf{159.17kg}$\pm$106.66kg   \\ \bottomrule
\end{tabular}

\vspace{1mm}
{\footnotesize Relative growth is normalized per mission by best achievable given $\trueCurrents$.}
\vspace{1mm}
\end{table}


\subsubsection{Case Study of 60-Day Scenario}
\begin{figure}[t]
    \vspace{2mm}
    \includegraphics[width=.46\textwidth,trim={0cm 0cm 0cm 1cm},clip]{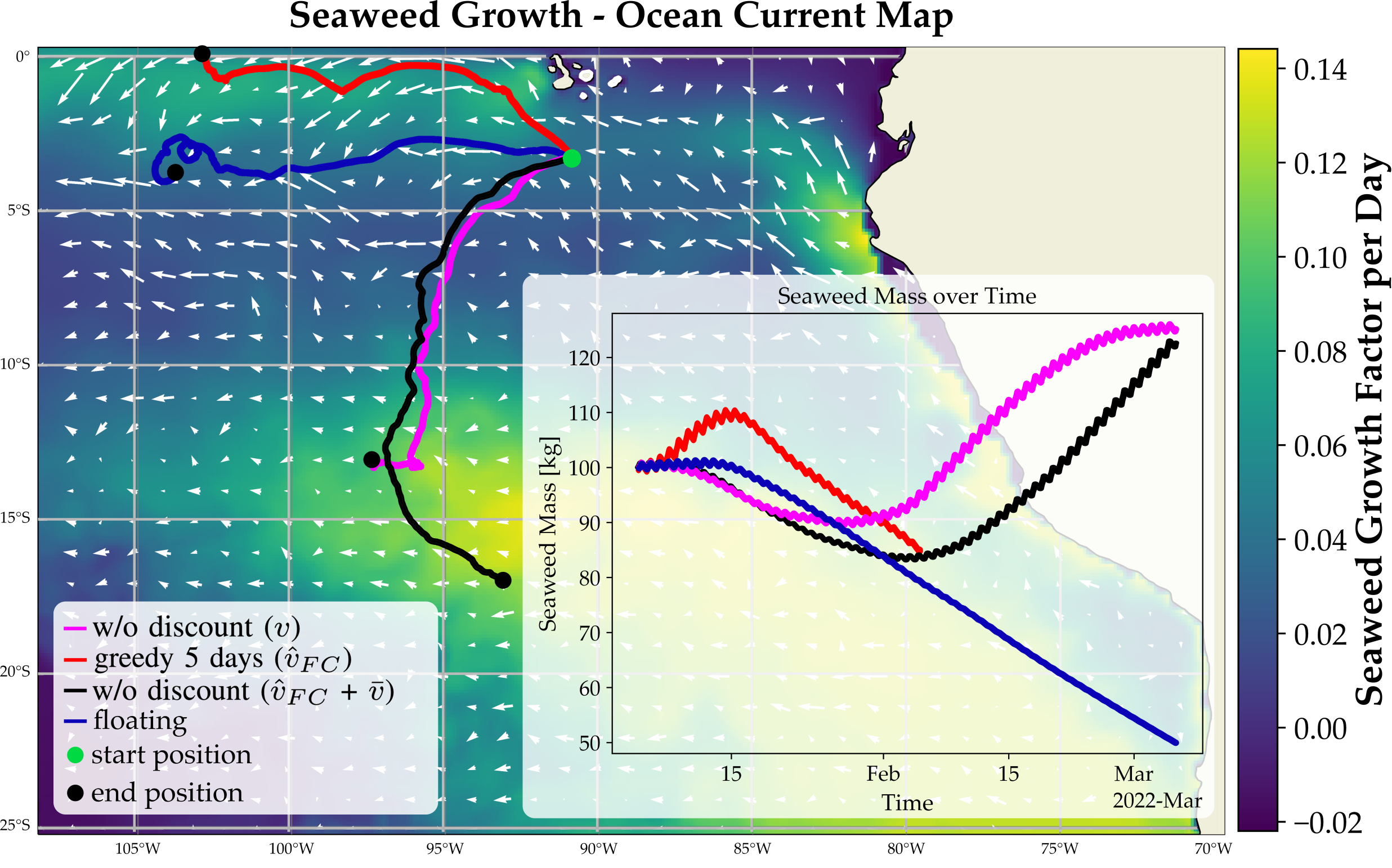}
    \caption{\small 60-day Case Study: The greedy controller optimizes for 5-day growth thereby navigating to the closest growth region. It fails to anticipate the strong currents that push it out of the region. The long-term controllers reach a more distant growth-richer area while incurring short-term losses.}
    \vspace{-2mm}
    \label{fig:showcase_tradeoff}
\end{figure}

We were intrigued that the 5-day controller did achieve almost the same seaweed growth as by planning over 30-days (Sec. \ref{subsubsec:exp_u_max_fixed}). Hence, we conducted a case study with planning and operating the farm over 60 days instead of 30 days and with $\bu_{max}\!\!=\!\!0.3 \frac{m}{s}$ (Fig. \ref{fig:showcase_tradeoff}). We find that 
the greedy controller then aims for the nearest growth region, while the long-term controller properly balances short-term lower growth against the long-term gains of reaching a high-growth region. This leads to the greedy controller being driven out of the simulated region while the long-term controller achieves close to the best achievable growth (see sub-figure Fig. \ref{fig:showcase_tradeoff}). Note that the zig-zags shape of the lines are due to day-night cycles.

\section{Discussion}
\label{sec:discussion}

\change{The experiments demonstrate that controllers using forecasts $\FCCurrents$ substantially outperform a passively floating farm. The myopic behavior of a greedy policy not only leads it to navigate toward low-growth regions in the vicinity but also fails to account for the possibility of being pushed out of good-growth regions by strong currents, as in our 60-day case study in Fig. \ref{fig:showcase_tradeoff}. Therefore, we were surprised that our 5-day optimizing controller was nearly on par with our 30-day optimizing controllers (Sect. \ref{subsubsec:exp_u_max_fixed}).

We attribute this to several factors. First, the initial position determines most of the possible growth within the 30 days Fig. \ref{fig:heatmap}. Farms starting from suboptimal positions cannot reach and grow seaweed in more distant high nutrient regions. We believe that experiments over the full seaweed growth cycle of 60-90 days would yield more significant differences between the controllers as long-term high growth and avoiding low growth regions becomes more important.} Second, the growth map in our region exhibits a smooth gradient, which means that even greedy controllers might move toward globally optimal growth regions without planning for it. Third, in our experimental evaluation, we do not consider missions where any controller leaves the predefined region (Sec. \ref{subsubsec:miss_gen}). This often occurs with greedy or floating controllers (Fig. \ref{fig:showcase_tradeoff}); consequently, the performance increase with long-term controllers would be greater if we accounted for the filtered missions.

\section{Conclusion and Future Work}
\label{sec:conclusion}
In this work, we maximize seaweed growth on autonomous farms that are underactuated and operate by harnessing uncertain ocean currents. We first introduced a 2D \ac{DP} formulation to solve for the growth-optimal value function when the true currents are known. Next, we showed how the value function computed on forecasted currents can be used as feedback policy for multiple farms, which is equivalent to replanning on the forecast at every time step and hence mitigates forecast errors. As operational forecasts are only 5 days long, we extended our method to reason beyond the forecast horizon by estimating expected future growth based on seasonal average currents. Lastly, we presented a finite-time discounting \ac{DP} \ac{PDE} to account for increasing uncertainty in ocean currents. 
We conducted extensive empirical evaluations based on realistic ocean conditions over 30 days. Our method achieved 95.8\% of the best achievable growth and 9.6\% more growth than passively floating despite its low propulsion of $u_{max}\!\!=\!\!0.1\frac{m}{s}$ and relying on daily 5-day forecasts. This demonstrates that low-power propulsion is a promising method to operate autonomous seaweed farms in real-world conditions.

A future direction is to learn the expected growth after the forecast horizon using experience and approximate value iteration \cite{beckenbach2022approximate} or a value network \cite{silver.2016}. This could implicitly learn the distribution shift between $\FCCurrents$ and $\trueCurrents$.
\mwnote{however, it may require intensive computation for training due to the necessity of i.i.d. samples, which could limit the number of samples taken per mission to just one \cite{silver.2016}.}
Another direction is to make the discount factor state-dependent based on the uncertainty of current predictions, which could be estimated historically or from forecast ensembles \cite{subramani_et_al_CMAME2018,lermusiaux_et_al_BBN_Oceans2020}. 
Lastly, we plan to conduct field tests with our partner \cite{phykos} to further validate our method in real-world ocean conditions.

\bibliographystyle{IEEEtran}
\bibliography{references,mseas}

\end{document}